\documentclass[11pt]{JHEP3}

\usepackage{epsfig}

\def\spose#1{\hbox to 0pt{#1\hss}}
\def\ltapprox{\mathrel{\spose{\lower 3pt\hbox{$\mathchar"218$}}
 \raise 2.0pt\hbox{$\mathchar"13C$}}}
\def\gtapprox{\mathrel{\spose{\lower 3pt\hbox{$\mathchar"218$}}
 \raise 2.0pt\hbox{$\mathchar"13E$}}}

\newcommand {\beq}{\begin{equation}}
\newcommand {\eeq}{\end{equation}}
\newcommand{\U}{\mathcal{U}}

\title{Matrix Models of Noncommutative (2d+1) Lattice Gauge Theories}
\author{Federica Bazzocchi, Michele Cirafici, Carlo Maccaferri and Stefano Profumo \\
        Scuola Internazionale Superiore di Studi Avanzati \\
	Via Beirut 2-4, I-34014 Trieste, Italy \\ 
	E-mail: \email{fbazzo@sissa.it, cirafici@sissa.it, maccafer@sissa.it, profumo@sissa.it} 
}


\preprint{}

\abstract{We investigate the problem of mapping, through the Morita equi\-va\-lence,
odd dimensional noncommutative lattice gauge theories onto suitable matrix
models. We specialize our analysis to noncommutative three dimensional QED (NCQED) and scalar QED (NCSQED), for which we
explicitly  build the corresponding Matrix Model.}

\newpage

\begin{document}

\section{Introduction}

The interest in noncommutative quantum field theories (see \cite{reviews} for a complete review of the subject) has been recently greatly enhanced by the discovery of new implications of the so called {\em Morita equivalence} \cite{Morita}. 
One important consequence of the Morita equivalence is in fact that U($N$) gauge theories with gauge fields obeying twisted boundary conditions over the noncommutative torus $\mathbb{T}^D_\Theta$ are equivalent to a U($\tilde{N}$) gauge theories, with $\tilde{N}$ suitably chosen, over the noncommutative torus $\mathbb{T}^D_{\Theta^\prime}$, with gauge fields obeying periodic boundary conditons. This algebraic relation has turned out to be very useful in order to promote a nonperturbative lattice study of noncommutative field theories. In particular, twisted Eguchi-Kawai (TEK) reduced models \cite{TEK} have been used as the commutative counterpart of noncommutative field theories \cite{AMNS1}, \cite{AMNS2}, \cite{amns} (for recent reviews of the subject see also \cite{Mak1}, \cite{Szab}, \cite{Mak2}). Among the field theories which have been studied so far there are the 2D Yang-Mills pure gauge theory on the lattice \cite{YM2}, the 2D principal chiral models \cite{profumo}, and the $\phi^4$ theory in 2D \cite{ambj}. One of the results which these studies have produced is a completely nonperturbative demonstration of the renormalizability of noncommutative field theories \cite{YM2, profumo}. This is mainly achieved by defining a {\em double scaling} limit to reproduce the continuum NC theory from its lattice definition and showing that Wilson loops \cite{YM2} or correlation functions \cite{profumo} indeed have finite limits when taking $N\rightarrow\infty$ and $a\rightarrow0$ keeping $Na^2\sim\theta$ constant, where $a$ is the lattice spacing and $\theta$ is the so called noncommutativity parameter. We recall that the TEK reduced models are equivalent to the corresponding non-reduced lattice field theories, through the correspondence of the respective sets of Schwinger-Dyson equations\footnote{The validity of the demonstration of the equivalence through the correspondence of the Schwinger-Dyson equations has been recently questioned in \cite{profumo2}.}, in the {\em planar limit}, i.e. taking $N\rightarrow\infty$ while keeping the lattice coupling constant fixed. The equivalence between NC lattice field theories and TEK matrix models is instead exact also at finite lattice size and matrix dimensions respectively \cite{reviews}. 
\\
The problems arising in the generalization of the above described procedure to an {\em odd number} of dimensions are two fold. On the one hand, in fact, matrix TEK models theories have been, in this context, introduced only in an even number of dimensions. On the other hand, until now the problem of encompassing fermions into TEK models has not been tackled in the context of the Morita equivalence. In fact, in \cite{fermMorita} the Morita equivalence for fermion theories on noncommutative two-tori has been studied, but without reference to TEK models, and once again in an even number of dimensions. Moreover, in \cite{DAS} it has been shown that in TEK models one can accommodate a number of fermion families which {\em must be} equal to the rank of the gauge group, i.e. if one has a SU($N$) gauge theory one can only accommodate $N_f$ fermion families in the reduced theory ({\em Veneziano limit}, $N,\ N_f\rightarrow\infty$ with $N/N_f=1$).\\
The solution we propose\footnote{While the present paper was in preparation, a similar idea has been suggested in \cite{simula} for the case of 3D NC $\phi^4$ theory on the lattice.} is based on a generalization of TEK models. Since in an odd number of dimensions one can always choose one of the coordinates (which we will call {\em time}) to commute with all the others, we can imagine to consider it as a standard lattice direction, i.e. without twisted boundary conditions. Thus we will have a {\em tower} of 2d-dimensional TEK models, one for each time slice. The action of the underlying NC theory will naturally generate interactions among next neighbours TEK models. We will show how this construction applies to pure gauge fields as well as to scalar and fermionic matter fields, in the simplest case of 3D. The procedure is nonetheless completely general, since the arbitrary even dimensional case has been exhaustively investigated in \cite{AMNS1}, \cite{AMNS2} and \cite{amns}.\\ 
As a test ground, we choose to investigate 3D NCQED. The choice is motivated by the possibility of applying the described procedure to any kind of NC field (scalars, fermions and gauge). Moreover, a lattice perturbation theory in noncommutative spaces has been recently formulated, and applied to 3D noncommutative QED (NCQED) \cite{QED}. In our paper we propose to investigate both the lattice 3D NCQED, and the 3D scalar noncommutative QED (NCSQED), via the above mentioned correspondence between noncommutative theories and matrix models.\\
The paper is organized as follows. We briefly set the notation in section \ref{NCfields}, describing in particular the NC space and fields we are dealing with. In section \ref{MM} we will explicitly construct our Matrix Model, and derive the expressions for the action of the matrix NCSQED and NCQED theories. Incidentally, we will generalize to the case of non abelian U($p$) gauge group for the noncommutative gauge field theory in subsection \ref{estensione}. Some concluding remarks are contained in section \ref{conclusions}.

\section{Noncommutative 3D quantum field theories on the lattice}\label{NCfields}

We consider 3D space-time with the standard commutation relation between the coordinates (Hermitian operators) $[ \hat{x}_\mu , \hat{x}_\nu ] = i \, \theta_{\mu\nu}$. We can choose, due to a property of antisymmetric matrices, the coordinates such that the third one (say euclidean {\em time}, after the Wick rotation) commutes with all the others. Therefore the noncommutativity anti-symmetric tensor can be set in the form
\begin{eqnarray}
\theta_{ij} \ \equiv 
\ \theta \    
\left(
\begin{array}{cccccc}
0 & & -1 &  & 0    \\
1 & &  0 &  & 0    \\
0 & &  0 &  & 0  
\end{array}
\right) \ .
\end{eqnarray}
The lattice regularization of our theory is given introducing a (2+1) dimensional toroidal lattice $\Lambda_{N,T}$ defined by
\begin{eqnarray}
\Lambda_{N,T}=\left\{ \ (x_{1},x_2,t) 
\in a \mathbb{Z}^{3} \ \Big|  \ 
a \leq  x_\mu 
\leq a {L_\mu} \ \right\} \ ,
\end{eqnarray}
where $a$ is the lattice spacing\footnote{The lattice spacing in the time direction can be choosen, in principle, to be different from the one in the space directions, but this is irrelevant for the following discussion and results.} and $L_1=L_2=N$, $L_3=T$.\\
The standard lattice star-product can be cast in the following integral form (see e.g. \cite{reviews}):
\beq
\phi_1(\vec{x},t) \star \phi_2(\vec{x},t)
= \frac{1}{N^{2}} \sum _{\vec{y} }
\sum _{\vec{z} }
 \phi_1(\vec{y},t) \, \phi_2 (\vec{z},t)
\, \mbox{e} ^{-2 \, i \, (\theta ^{-1})_{ij}  \, 
( x _i - y_i ) ( x _j - z_j )} \ ,
\eeq
where the noncommutativity parameter is taken to be
\beq
\theta  = 
\frac{1}{\pi} \, N \, a^2  \ .
\eeq

\subsection{The NC fields}

The U(1) NC gauge fields can be put on the lattice by \cite{QED}
\beq
\begin{array}{c}
\U_\mu (x) = {\cal P} \exp _{\star} \left( 
i g \int_x^{x +a \hat{\mu}} 
{\mathcal A}_\mu (s) \, \mbox{d} s \right) =\\
\\
= \sum_{n=0}^{\infty} \,  (ig)^n
\int_x^{x +a \hat{\mu}}  \mbox{d} \xi_1
\int_{\xi_1}^{x +a \hat{\mu}}  \mbox{d} \xi_2
\cdots 
\int_{\xi_{n-1}}^{x +a \hat{\mu}}  \mbox{d} \xi_n \ \ {\mathcal A}_\mu (\xi_1)  \star {\mathcal A}_\mu (\xi_2)  \star \cdots
\star {\mathcal A}_\mu (\xi_n)  \ ,\\
\end{array}
\eeq
where ${\mathcal A}_\mu(x)$ is the  
(real) gauge field in the continuum, and periodic boundary conditions are intended for the NC link fields in all lattice directions.\\
The NC link fields $\U_\mu(x)$ are {\em not} unitary, but {\em star-unitary},
\beq
\U_\mu(x) \star \U_\mu(x)^{*} = \U_\mu(x)^{*}  \star \U_\mu(x) = 1 \ .
\eeq
The continuum gauge field ${\mathcal A}_\mu (x)$ transforms
under the {\em star-gauge transformation} as
\beq
{\mathcal A}_{\mu}(x)  \mapsto  g(x) 
\star {\mathcal A}_{\mu}(x) \star g(x)^{\dagger}
- \frac{i}{g} g(x) \star \frac{\partial}{\partial x_{\mu}} 
g(x)^{\dagger} \ ,
\eeq
where $g(x)$ is also star-unitary.
Under this transformation, the link field $\U_\mu(x)$ 
transforms as
\begin{eqnarray}
\U_{\mu}(x) \mapsto g(x) \star \U_{\mu}(x) \star 
g(x+a\hat{\mu}) ^{*} \ .
\label{trans_gauge}
\end{eqnarray}
As far as the matter fields are concerned, we will consider NC scalar fields $\varphi(x)$ and fermionic fields $\psi(x)$. We will impose periodic boundary conditions for the scalar fields, while we will discuss in section \ref{fermioncini} the slightly more subtle case of fermion fields.\\
In the case the matter fields transform in the fundamental representation, they will satisfy the following star-gauge transformations
\beq
\varphi(x) \mapsto g(x)\star \varphi(x); \hspace*{1cm} 
\varphi(x)^* \mapsto \varphi(x)^*\star g(x)^{*} 
\eeq
and
\beq
\psi(x) \mapsto g(x)\star \psi(x); \hspace*{1cm} 
\bar{\psi}(x) \mapsto \bar{\psi}(x)\star g(x)^{*}. 
\eeq
For the adjoint representation we have instead respectively
\beq
\varphi(x)\mapsto g(x)\star\varphi(x)\star g(x)^{*}; 
\hspace*{1cm} \varphi(x)^*\mapsto g(x)\star \varphi(x)^*\star 
g(x)^{*} 
\eeq
and
\beq
\psi(x)\mapsto g(x)\star\psi(x)\star g(x)^{*}; 
\hspace*{1cm} \bar{\psi}(x)\mapsto g(x)\star \bar{\psi}(x)\star 
g(x)^{*} \ .
\eeq

\subsection{The NCSQED and NCQED lattice actions}

The lattice action for the NCSQED (NCQED) is given by the sum of the gauge fields action and of the scalar (fermionic) action: 
\begin{eqnarray}
S=S_{\rm G}+S_{\rm S}\qquad (\mbox{NCSQED})\\
S=S_{\rm G}+S_{\rm F}\qquad (\mbox{NCQED})
\end{eqnarray}
which are respectively defined by
\begin{eqnarray}
S_{\rm G} = -\beta 
\sum_{x\in \Lambda_{T,N}}\sum_{\mu\neq\nu} 
\U_{\mu}(x)\star \U_{\nu}(x+a \hat{\mu})
\star \U_{\mu}(x+a \hat{\nu})^{*}\star 
\U_{\nu}(x)^{*} \ ,
\label{action}
\end{eqnarray}
by
\begin{eqnarray}
S_{\rm S}=
a^3\sum_{x}\Big[\left(\nabla_\mu\varphi(x)^*\right)\star\left(\nabla_\mu\varphi(x)\right)+m^2_{\varphi}\varphi(x)^*\star\varphi(x)\Big]
\label{scalaract}
\end{eqnarray}
and by
\begin{eqnarray}
S_{\rm F}=
a^3 \sum_{x}\bar{\psi}(x)\star (D_{\rm w}-m_\psi) \, \psi(x) \ ,
\label{feract}
\end{eqnarray}
where $D_{\rm w}$ is the Dirac-Wilson operator
\begin{eqnarray}
D_{\rm w} = {1\over 2}\sum_{\mu=1}^{3}
\Big[\gamma_{\mu}(\nabla^{*}_{\mu}+\nabla_{\mu})
+ra \nabla^{*}_{\mu}\nabla_{\mu}\Big] \ .
\label{DW}
\end{eqnarray}
The second term in (\ref{DW}) is the Wilson term, which gives to the species doublers a mass of $\mathcal{O}(1/a)$. The coefficient $r$ can be taken to be unity, although it can take other values as far as its magnitude is of order one \cite{QED}.\\
The expression of the forward and backward 
covariant derivatives depends on the transformation properties of the
fermionic (scalar) field. In the case where $\varphi(x)$ (or $\psi(x)$) transforms 
in the fundamental representation they are given respectively by
\begin{eqnarray}
\nabla_{\mu}\varphi &=& {1\over a}
\left[\U_{\mu}(x)\star\varphi(x+a\hat{\mu})-\varphi(x)\right] \nonumber \\
\nabla_{\mu}^{*}\varphi &=& {1\over a}\left[\varphi(x)-
\U_{\mu}(x-a\hat{\mu})^{*}
\star \varphi(x-a\hat{\mu})\right] \ ,
\label{covder_fund}
\end{eqnarray}
and analogously for the fermionic case. On the other hand, when scalars or fermions transform in the adjoint representation the forward and backward covariant derivatives are respectively defined by
\begin{eqnarray}
\nabla_{\mu}\varphi &=& 
{1\over a}\left[\U_{\mu}(x)\star\varphi(x+a\hat{\mu})
\star \U_{\mu}(x)^{*}
-\varphi(x)\right] \nonumber \\
\nabla_{\mu}^{*}\varphi &=& {1\over a}
\left[\varphi(x)-\U_{\mu}(x-a\hat{\mu})^* 
\star \varphi(x-a\hat{\mu})
\star \U_{\mu}(x-a\hat{\mu}) \right] \ .
\label{covder_adj}
\end{eqnarray}
In either case, the scalar and fermion actions (\ref{scalaract}) and (\ref{feract}) are
star-gauge invariant.

\section{The Morita-equivalent Matrix Model}\label{MM}

The idea we propose is to build a tower of $T$ TEK models, one for each time slice of the lattice. In this way, the lattice traslations in the NC directions 1 and 2 are implemented via the so called twist eaters $\Gamma_{1,2}$ , which act as lattice shift operators for the (matrix) fields obeying twisted boundary conditions. On the other hand, in the action there will be an interaction term between the time next-neghbour TEK-like reduced models, and therefore the theory will differ from the standard, and well known, lattice gauge reduced model.\\
We define, as usual, the shift and clock twist eaters, obeying the Weyl-'t Hooft algebra \mbox{$\Gamma_i\Gamma_j=\exp\Big[\frac{2\pi i}{N}\varepsilon_{ij}\Big]\Gamma_j\Gamma_i$} (where $\varepsilon_{ij}$ is the totally anti-symmetric 2D tensor):
\begin{eqnarray}
\Gamma_1 \ =  \ \left(
\begin{array}{cccccc}
0 & 1 &  &  & 0    \\
  & 0 & 1 &  &     \\
  & & \ddots & \ddots & \\
  & & & \ddots & 1\\
1 &&&& 0  
\end{array}
\right)
\label{gamma1}
\end{eqnarray}
\begin{eqnarray}
\Gamma_2 \ =  \ \left(
\begin{array}{cccccc}
1 &  &  &  &     \\
  & \mbox{e}^{2\pi i/N} &  &  &     \\
  & & \mbox{e}^{4\pi i/N} & & \\
  & & & \ddots & \\
  &&&& \mbox{e}^{2(N-1)\pi i/N}  
\end{array}
\right)
\label{gamma2}
\end{eqnarray}
and the following set of $N \times N$ unitary unimodular matrices
\beq
J_k=\Gamma_1^{k_1}\Gamma_2^{k_2}\,\mbox{e}^{\pi i k_1k_2/N}.
\label{Jkdef}
\eeq
The $J_k$ are defined for integer valued vectors $k$.\\
Since
$(\Gamma_i)^N=1_N$, these matrices have the periodicity properties
\beq
J_{N-k}=J_{-k}=J_k^\dagger \ ,
\label{Jkperiod}\eeq
and they obey the algebraic relations
\beq
J_k\,J_q=\prod_{i=1}^N\,\prod_{j=1}^N\mbox{e}^{\pi ik_i\varepsilon_{ij}q_j/N}~J_{k+q} \ .
\eeq
The $J_k$'s have the same formal algebraic properties as the plane wave Weyl
basis $\mbox{e}^{ik_i\hat x^i}$ for the continuum noncommutative field theory on the
torus. Owing to the
property (\ref{Jkperiod}), there are only $N^2$ independent matrices.\\
The matrices (\ref{Jkdef}) obey the orthonormality and completeness relations
\begin{eqnarray}
\frac1N\,\mbox{tr}_N\left(J_k\,J_q^\dagger\right)&=&\delta^{(2)}_{k,q\,({\rm mod}
\,N)} \ , \\
\frac1N\,\sum_{k\in{\mathbb Z}_L^D}\,\left(J_k\right)_{ab}\,\left(J_k
\right)_{cd}&=&\delta_{ad}\,\delta_{bc} \ .
\end{eqnarray}
They thereby form the {\it Weyl basis} for the linear space $gl(N,{\mathbb C})$ of
$N\times N$ complex matrices.\\ 
In complete analogy with the continuum formalism, we define
these via the $N\times N$ matrix fields
\beq
\Delta(\vec{x})=\frac1{N^2}\,\sum_{k\in{\mathbb Z}_N^2}J_k~\mbox{e}^{-2\pi ik_ix^i/l} \ ,
\label{Deltafinite}\eeq
where
\beq
l=a N
\eeq
is the dimensionful extent of the hypercubic lattice with $N^2$ sites
$x^i$. Because of the relations (\ref{Jkperiod}), the matrices $\Delta(x)$ are
Hermitian and periodic in $x^i$ with period $l$. This means that the
underlying lattice is a discrete torus.  Among the algebraic relations satisfied
by the matrices $J_k$ and by the operator $\Delta(\vec{x})$ there are the following relations
\beq
\mbox{tr}_N\Bigl(J_k\,\Delta(\vec{x})\Bigr)\ =\ \frac1N~\mbox{e}^{2\pi ik_ix^i/l} \ , 
\label{Jkalgebraicconditions1}
\eeq
\beq
\frac1N\,\sum_x\,\Delta(\vec{x})_{ab}\,\Delta(\vec{x})_{cd}\ =\ \delta_{ad}
\,\delta_{bc}\ ,
\eeq
\beq
\frac1N\,\mbox{tr}_N\Bigl(\Delta(\vec{x})\,
\Delta(\vec{y})\Bigr)\ =\ N^2\,\delta^{(2)}_{\vec{x},\vec{y}\,({\rm mod}\,l)} \ .
\label{Jkalgebraicconditions}
\eeq

\subsection{The Matrix Model (I): pure gauge action}\label{puregauge}

The matrix theory we propose is given by $T$ lattice sites. At each lattice site, say $t$, there is a $1^2$ sublattice in the links of which live the unitary $N\times N$ matrices $U_1(t)$ and  $U_2(t)$. In the link between lattice site $t$ and $t+1$ lives the $N\times N$ unitary matrix $U_3(t)$.\\
In what follows we will indicate with Greek letters ($\mu$) the indices running from 1 to 3, and with Latin indices those running only on space coordinates ($i=1,2$).\\ 
We can in fact make a discrete Fourier transformation in the space coordinates to define lattice fields on a
discrete torus,
\beq
U_\mu(t)=\frac1{N^2}\,\sum_{k\in \mathbb{Z}_N^2}U_\mu(k,t)\ J_k\quad ,\quad
U_\mu(k,t)=N\ \mbox{tr}_N\left(U_\mu(t)\ J_k^\dagger\right) \ .
\label{Uikdef}\eeq
We then associate the lattice NC link fields  $\U_{\mu} (\vec{x},t)$ to the matrix variables by the Fourier series
\beq
{\U}_\mu(\vec{x},t)\equiv\frac1{N^2}\,
\sum_{k\in \mathbb{Z}_N^2}U_\mu(k,t)~\mbox{e}^{2\pi ik_ix^i/l}=\frac1N\,
\mbox{tr}_N\Bigl(U_\mu(t)\,\Delta(\vec{x})\Bigr) \ .
\label{calUix}
\eeq  
The star-unitarity condition on the NC link fields $\U_{\mu} (\vec{x},t)$
\beq
{\cal U}_\mu(\vec{x},t)\star{\cal U}_\mu(\vec{x},t)^*={\cal U}_\mu(\vec{x},t)^*\star{\cal U}_\mu(\vec{x},t)=1 \ ,
\label{latticestarunitary}\eeq
implies, on the $N\times N$ matrix variables
\beq
U_\mu(t)=\frac1{N^2}\,\sum_{\vec{x}}\,{\cal U}_\mu(\vec{x},t)\,\Delta(\vec{x}),
\label{UiDelta}\eeq
the standard unitarity condition
$U_\mu(t)\,U_\mu(t)^\dagger=U_\mu(t)^\dagger\,U_\mu(t)=1_N\quad \forall t$.\\ 
Using the definition of the $\Gamma_i$ matrices, and equations (\ref{Jkdef}) and (\ref{Deltafinite}) one can easily calculate
\beq
\Gamma_i\,\Delta(\vec{x})\,\Gamma_i^\dagger=\Delta(\vec{x}-a\,\hat\imath) \ , \quad i=1,2,
\label{GammaDeltashift}\eeq
from which it follows that spatial shifts of the lattice NC gauge fields may be
represented as
\beq
\U_\mu(\vec{x}+a\,\hat\imath,t)=\frac1N\,\mbox{tr}_N\Bigl(\Gamma_i\,U_\mu(t)\,
\Gamma_i^\dagger\,\Delta(\vec{x})\Bigr) \ .
\label{Uishift}\eeq
Using the completeness relation
\beq
\frac1{N^2}\,\sum_x\,\Delta(\vec{x})=\mathbb{I}_N
\label{completeness}\eeq
we can substitute into (\ref{action}) the definition (\ref{calUix}) obtaining, from the sum over different directions, a part $\{1,2\}$, $\{2,1\}$ which reproduces, for each lattice slice $t$, a TEK model, and four mixed parts, $\{1,3\}$, $\{3,1\}$, $\{2,3\}$ and $\{3,2\}$.\\
Using (\ref{Uishift}) and (\ref{Jkalgebraicconditions}) it is straightforward to write the pure gauge action 
\beq
\begin{array}{rcl}
S_{\rm G}^{N\times N} & = & -\beta \sum_{t=1}^T \mbox{tr}_N\Bigl( A_{TEK}(t) \ +  \\
&&\\
 & & U_1(t) \ \Gamma_1 \  U_3(t) \ \Gamma^\dagger_1  \ U_1^\dagger (t+1)  \ U_3^\dagger(t) \ + \ \mbox{h.\ c.}\\
&&\\
& & U_2(t) \ \Gamma_2 \ U_3(t) \ \Gamma^\dagger_2 \ U_2^\dagger (t+1) \ U_3^\dagger(t) \ + \ \mbox{h.\ c.}\ \ \Bigr),\\
\end{array}
\label{actionNN}
\eeq
where 
\beq
A_{TEK}(t) \ = \ U_1(t) \ \Gamma_1 \ U_2(t)\ \Gamma_1^\dagger\  \Gamma_2\  U_1^\dagger(t)\  
\Gamma_2^\dagger\ U_2^\dagger(t)\ + \ \mbox{h.\ c.}
\label{azionetekke}
\eeq
Equation (\ref{azionetekke}) reproduces the well known TEK action \mbox{$S_{TEK}=\sum_{i\neq j}\mathcal{Z}^*_{ij}\ \mbox{tr}_N\left(V_i V_j V^\dagger_i V^\dagger_j\right)$}, where \mbox{$V_i\equiv U_i\Gamma_i$} and $\mathcal{Z}_{ij}\equiv\exp\Big[\frac{2\pi i}{N}\varepsilon_{ij}\Big]$.\\
Eq. (\ref{actionNN}) completes the definition of the $N\times N$ matrix action as far as the gauge fields part is concerned.

\subsubsection{Generalization to NC U($p$) Yang-Mills theories}\label{estensione}

The above outlined construction can be staightforwardly generalized to the case where the gauge group of the noncommutative field theory is non-abelian. If we consider the gauge groups U($p$) for the noncommutative theory, we have to use $Np\times Np$ unitary U($Np$) matrices, and take the trace $\mbox{Tr}_p$ on the subgroup SU($p$) of U($Np$) \cite{amns}.\\ First, we have to re-define the $\Gamma_i$ matrices of eq. (\ref{gamma1}) and (\ref{gamma2}) as the following tensor products
\beq
\Gamma^p_i\ \equiv \ \Gamma_i\otimes\ \mathbb{I}_p.
\label{newgammas}
\eeq
The NC link fields $\U^p_i(\vec{x},t)$ as functions of the new matrix variables $U^p_i(t)$ will then simply read
\beq
{\U}^p_i(\vec{x},t)\equiv\frac1N\, \mbox{Tr}_p\Big[
\mbox{tr}_N\Bigl(U^p_i(t)\,\Delta^p(\vec{x})\Bigr)\Big] \ ,
\eeq  
where the new $\Gamma^p_i$ entering in the definition of $\Delta^p(x)$ of eq. (\ref{Deltafinite}) are the one of eq. (\ref{newgammas}). The lattice shifts will correspondingly be implemented by these new matrices, and the non-abelian action will differ from (\ref{actionNN}) only for the extra  $\mbox{Tr}_p$ on the subgroup SU($p$).

\subsection{The Matrix Model (II): scalar action}

In analogy with what has been done for the gauge link fields, it is straightforward to use the $J_k$'s basis of the U(N) algebra to map the noncommutative $\varphi(\vec{x},t)$ scalar fields onto the $N\times N$ matrix variables $\phi(t)$, this time defined on the {\em lattice sites} (and no longer on the links) through the Fourier transform of the noncommutative fields.\\
In fact, we perform, exactly as above, a discrete Fourier transformation to define
\beq
\phi(k,t)=N\,\mbox{tr}_N\left(\phi(t)\,J_k^\dagger\right)
\eeq
and then the noncommutative variables arise as
\beq
\varphi(\vec{x},t)\equiv\frac1{N^2}\,
\sum_{k\in \mathbb{Z}_N^2}\phi(k,t)~\mbox{e}^{2\pi ik_ix^i/l}=\frac1N\,
\mbox{tr}_N\Bigl(\phi(t)\,\Delta(\vec{x})\Bigr) \ .
\eeq
Thanks to eq. (\ref{GammaDeltashift}), the spatial shifts of the lattice NC fields are represented by
\beq
\varphi(\vec{x}+a\,\hat\jmath,t)=\frac1N\,\mbox{tr}_N\Bigl(\Gamma_j\,\phi(t)\,
\Gamma_j^\dagger\,\Delta(\vec{x})\Bigr) \ .
\eeq 
We now proceed to the calclulation of the scalar action $S_{\rm S}$.\\
In the case of the {\em fundamental representation} we have:
\beq
S^{N\times N}_{\rm S}\ =\ a\ \sum_{i=1,2}S^i_{\rm S}+S^3_{\rm S}\ +\ a^3\ m_{\phi^2}^2\sum_{t=1}^T\mbox{tr}_N\left(\phi^\dagger(t)\ \phi(t)\right)
\eeq
where
\beq
S^i_{\rm S}\ = \ \sum_{t=1}^T\mbox{tr}_N\Big[\phi^\dagger(t)\ \phi(t)-\Gamma_i\ \phi^\dagger(t)\ \Gamma_i^\dagger\ U_i^\dagger(t)\ \phi(t)\ +\ \mbox{h.\ c.}\ \Big].
\eeq
and
\beq
S^3_{\rm S}\  = \ \sum_{t=1}^T\mbox{tr}_N\Big[\phi^\dagger(t)\ \phi(t)-\ \phi^\dagger(t+1)\  U_3^\dagger(t)\ \phi(t)\ +\ \mbox{h.\ c.}\ \Big],
\eeq
In the case of the {\em adjoint representation} we have instead
\beq
S^i_{\rm S}\ =
\ \sum_{t=1}^T\mbox{tr}_N\Big[\phi^\dagger(t)\ \phi(t)-\ U_i(t)\ \Gamma_i\ \phi(t)\ \Gamma_i^\dagger\ U_i^\dagger(t)\ \phi^\dagger(t)\ +\ \mbox{h.\ c.}\ \Big],
\eeq
and
\beq
S^3_{\rm S}\ = \ \sum_{t=1}^T\mbox{tr}_N\Big[\phi^\dagger(t)\ \phi(t)-\ U_3(t)\ \phi(t+1)\  U_3^\dagger(t)\ \phi^\dagger(t)\ +\ \mbox{h.\ c.}\ \Big].
\eeq
These expressions complete the definition of the matrix model of NCSQED, which overall consists, given $T$ time-like lattice sites, of $3T$ unitary matrices $U_i(t)$ and $T$ $N\times N$ $\phi(t)$ matrices. We recall that the limit $T\rightarrow\infty$ can be taken as in commutative theories, and one can have arbitrary lenght $\tau=Ta$ indipendently of $\theta$ and $l=Na$.\\

\subsection{The Matrix Model (III): fermionic action}\label{fermioncini}

We can have four different spin structures for the fermionic fields on the lattice. We have in fact the following boundary conditions (we recall that $l=aN$ and $\tau=aT$):
\beq
\begin{array}{rcl}
\psi(x_1+l,x_2,t)\ &=\ &\mbox{e}^{2\pi i \alpha_1}\psi(x_1,x_2,t),\\
&&\\
\psi(x_1,x_2+l,t)\ &=\ &\mbox{e}^{2\pi i \alpha_2}\psi(x_1,x_2,t), \\
&&\\
\psi(x_1,x_2,t+\tau)\ &=\ &-\psi(x_1,x_2,t),
\end{array}
\label{fermbound}
\eeq
where $\alpha_1$ and $\alpha_2$ can take the values 0,1/2. In order to formulate a finite temperature lattice field theory on the torus, the boundary condition in the time direction has to be taken anti-periodic for fermions. We will call a fermion with boundary conditions (\ref{fermbound}) as of type $\vec{\alpha}=(\alpha_1,\alpha_2)$ and denote it $\psi_{\vec{\alpha}}$.\\
We will show in what follows how to map fermions with any spin structure into the Matrix Model under study. Exactly as above, we have to map the algebra of the NC fermionic fields into the U($N$) algebra. This will be done through a correspondence between the Fourier modes of the NC fermionic fields and, once again, the basis of the U($N$) algebra given by the matrices of eq. (\ref{Jkdef}).\\
The Fourier expansion in the space coordinates of a fermion field of type $\vec{\alpha}$ has the following form \cite{fermMorita}:
\beq
\psi_{\vec{\alpha}}(\vec{x},t)\ =\ \frac{1}{N^2}\sum_{k\in {\mathbb Z}^2_N}\psi(k,t)U_{\vec{k}+\vec{\alpha}},\quad \mbox{with} \quad U_{\vec{k}}\equiv\exp\left(2\pi i \vec{k}\cdot\vec{x}/l\right)
\eeq
The straighforward computation of the star-commutators of the Fourier basis $U_{\vec{k}}$ gives
\begin{equation}
[U_{\vec{k}+\vec{\alpha}},U_{\vec{k}'+\vec{\alpha}'}]_\star=
-2i\sin\left(\frac{\pi}{N}(2\vec{k} + 2\vec{\alpha})
\wedge(2\vec{k}^\prime + 2\vec{\alpha}^\prime)\right)U_{\vec{k} +
\vec{k}'+\vec{\alpha} + \vec{\alpha}'}, \label{algebra}
\end{equation}
where $\vec{\alpha}\wedge\vec{\beta}=\varepsilon_{ij}\alpha_i\beta_j$.
The algebra (\ref{algebra}) is isomorphic to the
 Lie algebra U($N$). A N-dimensional representation
of this algebra can be constructed with the shift and clock matrices of eq. (\ref{gamma1}) and (\ref{gamma2}). The matrices $J_k$ of eq. (\ref{Jkdef}) generate an algebra isomorphic to (\ref{algebra}). In fact we can rewrite the algebra of the $J_k$'s as \cite{fermMorita}
\begin{equation}
 [J_{\vec{n}},J_{\vec{m}}]=
 - 2i \sin\left(\frac{\pi}{N}\vec{n}
 \wedge\vec{m}\right)J_{\vec{n}+\vec{m}}.
\end{equation}
Therefore one has a Morita mapping between the Fourier modes defined on a lattice on the noncommutative torus and functions taking values on
U($N$), on a commutative space:
\begin{equation}
\exp\left(2\pi i(\vec{k}+\vec{\alpha})\cdot
\hat{\vec{x}}/l)\right)\, \leftrightarrow \,\exp\left(2\pi
i(\vec{k}+\vec{\alpha})\cdot
\vec{x}/l\right)J_{2(\vec{k}+\vec{\alpha})} \label{morita}
\end{equation}
This correspondence generates a mapping between fermion fields in the
following way: using (\ref{morita}), we define the Morita
equivalent fermion fields as
\begin{equation}
\Psi(t)=\frac{1}{N^2}\sum_{k\in{\mathbb Z}^2_N}\psi(k,t)J_{\vec{k}} \label{map1}
\end{equation}
As far as the Dirac structure of the spinors is concerned we consider a hermitian representation of the $2\times 2$ $\gamma$ matrices satisfying $\gamma_\mu\gamma_\nu=\delta_{\mu\nu}+i\epsilon_{\mu\nu\sigma}\gamma_\sigma$. The Pauli matrices provide an example of such a representation. We recall also that in odd dimensions there are no chiral fermions. The $N\times N$ matrix variables form therefore now a Dirac spinor $\left(\Psi_1(t),\Psi_2(t)\right)$, which once again will be defined on the $T$ lattice sites of the time direction.\\
In our definition (\ref{map1}) the four different spin structures are encompassed in the matrix spinors $\Psi(t)$. In order to define the correspondence between $\Psi(t)$ and the NC fields $\psi_\alpha(t)$, we need the fermion analogue of (\ref{Deltafinite}), which depends on the spin structure one has. We will therefore define
\beq
\Delta^{\vec{\alpha}}(\vec{x})=\frac1{N^2}\,\sum_{k\in{\mathbb Z}_{N}^2}J_{2\vec{k}+2\vec{\alpha}}~\mbox{e}^{-2\pi i(k_i+\alpha_i)x^i/l}.
\eeq
Due to the orthonormality condition (\ref{Jkalgebraicconditions1}) given a spin structure $\vec{\alpha}$ it is possible to select from the matrix variable $\Psi(t)$ (which is blind to it) the original noncommutative field (which is {\em not} blind to it) simply by
\beq
\psi_{\vec{\alpha}}(\vec{x},t)\ =\  \frac{1}{N}\ \mbox{tr}_N\Bigl(\Psi(t)\,\Delta^{\vec{\alpha}}(\vec{x})\Bigr).
\eeq
Moreover, we will have in the noncommutative planes at constant $t$
\beq
\psi_{\vec{\alpha}}(\vec{x}+a\,\hat\jmath,t)=\frac{1}{N}\,\mbox{tr}_{N}\Bigl(\Gamma_j\,\Psi(t)\,
\Gamma_j^\dagger\,\Delta^{\vec{\alpha}}(\vec{x})\Bigr) \ .
\label{psishift}
\eeq
By virtue of (\ref{psishift}) we can rewrite the matrix model action for the fermions straightforwardly, setting, as in the case of the scalars,
\beq
S^{N\times N}_{\rm F}=\sum_{i=1}^2S^i_{\rm F}+S^3_{\rm F}\ -\  a^3\ m_\psi\ \sum_{t=1}^T \mbox{tr}_N\left(\overline{\Psi}(t)\Psi(t)\right),
\eeq
and using (\ref{psishift}) we get, for the {\em fundamental representation}
\beq
\begin{array}{rcl}
S^i_{\rm F} & = & \frac{a^2}{2}\sum_{t=1}^T\mbox{tr}_N\Big[\overline{\Psi}(t)\Big[\gamma_i\left(U_i(t)\Gamma_i\Psi(t)\Gamma_i^\dagger-\Gamma_i^\dagger U_i^\dagger(t)\Psi(t)\Gamma_i\right)+\\
&&\\
&& + r \left(U_i(t)\Gamma_i\Psi(t)\Gamma_i^\dagger+\Gamma_i^\dagger U_i^\dagger(t)\Psi(t)\Gamma_i-2\Psi(t)\right)\Big]\Big]
\end{array}
\eeq
and 
\beq
\begin{array}{rcl}
S^3_{\rm F} & = & \frac{a^2}{2}\sum_{t=1}^T\mbox{tr}_N\Big[\overline{\Psi}(t)\Big[\gamma_3\left(U_3(t)\Psi(t+1)- U_3^\dagger(t-1)\Psi(t-1)\right)+\\
&&\\
&& + r \left(U_3(t)\Psi(t+1)+ U_3^\dagger(t-1)\Psi(t-1)-2\Psi(t)\right)\Big]\Big].
\end{array}
\eeq
Instead we find for the {\em adjoint representation}
\beq
\begin{array}{rcl}
S^i_{\rm F} & = & \frac{a^2}{2}\sum_{t=1}^T\mbox{tr}_N\Big[\overline{\Psi}(t)\Big[\gamma_i\left(U_i(t)\Gamma_i\Psi(t)\Gamma_i^\dagger U_i^\dagger(t)-\Gamma_i^\dagger U_i^\dagger(t)\Psi(t)U_i(t)\Gamma_i\right)+\\
&&\\
&& + r \left(U_i(t)\Gamma_i\Psi(t)\Gamma_i^\dagger U_i^\dagger(t)+\Gamma_i^\dagger U_i^\dagger(t)\Psi(t)U_i(t)\Gamma_i-2\Psi(t)\right)\Big]\Big]
\end{array}
\eeq
and 
\beq
\begin{array}{rcl}
S^3_{\rm F} & = & \frac{a^2}{2}\sum_{t=1}^T\mbox{tr}_N\Big[\overline{\Psi}(t)\Big[\gamma_3\left(U_3(t)\Psi(t+1)U_3^\dagger(t)- U_3^\dagger(t-1)\Psi(t-1)U_3(t-1)\right)+\\
&&\\
&& + r \left(U_3(t)\Psi(t+1)U_3^\dagger(t)+U_3^\dagger(t-1)\Psi(t-1)U_3(t-1)-2\Psi(t)\right)\Big]\Big].
\end{array}
\eeq
These expressions, together with the ones obtained in section \ref{puregauge}, complete the definition of the full lattice NCQED in 3D as an explicit matrix theory.

\section{Concluding remarks}\label{conclusions}

We showed how to generalize the correspondence between noncommutative lattice field theories and $N\times N$ Matrix Models to an arbitrary (odd) number of dimensions. In particular, we have explicitly built one of such Matrix Models for 3D noncommutative scalar and fermion QED.\\
It would be of great interest to address, in the present setting, the issue of NC Chern-Simons theory \cite{ChernSimons} on the lattice \cite{ChernSimons2}.\\
The explicit matrix formulation of NC lattice field theories in any number of dimensions could also be regarded as a valuable tool for numerical studies. The intriguing open questions of the nonperturbative renormalizability of NCQED and NCSQED, as well as the possible existence of a QED confinement-deconfinement transition \cite{confin} in the noncommutative case, are among the further possible developments of this work.\\

\acknowledgments{We are very grateful to Elisabetta Pallante for her help, valuable comments and remarks. We acknowledge Loriano Bonora for illuminating suggestions. We also thank Fernando Alday and Martin O'Laughlin for discussions. One of us (S.P.) thanks Jun Nishimura and Miguel Vazquez-Mozo for clarifications about \cite{QED}.}


\end{document}